\documentclass[usenatbib,useAMS]{mn2e}

\usepackage{wasysym}
\usepackage{graphicx}
\usepackage{longtable}
\usepackage{caption}
 
\begin{document}

\label{firstpage}

\title[Origin of dwarf galaxies]{Dwarf elliptical galaxies as ancient tidal dwarf galaxies}

\author[Dabringhausen \& Kroupa]{
J. Dabringhausen$^{1}$ \thanks{E-mail: joedab@astro.uni-bonn.de},
P. Kroupa$^{1}$ \thanks{pavel@astro.uni-bonn.de}\\
$^{1}$ Argelander Institute f\"ur Astronomie, Universit\"at Bonn, Auf dem H\"ugel 71, 53121 Bonn, Germany}

\pagerange{\pageref{firstpage}--\pageref{lastpage}} \pubyear{2012}

\maketitle

\begin{abstract}
The formation of tidal dwarf galaxies (TDGs) is triggered by the encounters of already existing galaxies. Their existence is predicted from numerical calculations of encountering galaxies and is also well documented with observations. The numerical calculations on the formation of TDGs furthermore predict that TDGs cannot contain significant amounts of non-baryonic dark matter. In this paper, the first exhaustive sample of TDG-candidates from observations and numerical calculations is gathered from the literature. These stellar systems are gas-rich at the present, but they will probably evolve into gas-poor objects that are indistinguishable from old dwarf elliptical galaxies (dEs) based on their masses and radii. Indeed, known gas-poor TDGs appear as normal dEs. According to the currently prevailing cosmological paradigm, there should also be a population of primordial galaxies that formed within haloes of dark matter in the same mass range. Due to their different composition and origin, it would be expected that objects belonging to that population would have a different structure than TDGs and would thus be distinguishable from them, but such a population cannot be identified from their masses and radii. Moreover, long-lived TDGs could indeed be numerous enough to account for all dEs in the Universe. Downsizing, i.e. that less massive galaxies tend to be younger, would then be a natural consequence of the nature of the dEs. If these claims can be kept up in the light of future observations, the presently prevailing understanding of galaxy formation would need to be revised.
\end{abstract}

\begin{keywords}
galaxies: dwarf -- galaxies: evolution -- galaxies: formation -- dark matter
\end{keywords}

\section{Introduction}
\label{sec:Introduction}
 
Observations show that encountering galaxies often have bridges of matter connecting them or elongated arcs of matter extending from them \citep{Zwicky1956}. Well known examples are the Antennae Galaxies (NGC 4038 and NGC 4039) and the Mice Galaxies (NGC 4676A and NGC 4676B). Theoretically, the formation of these filamentary structures can be understood by gravitational forces that the encountering galaxies exert on each other \citep{Toomre1972}. These gravitational forces lead to a distortion of the galaxies, because strength and direction of an external gravitational force depends on the location within a galaxy. The position-dependent changes of the external force within the galaxy are known as tidal forces, and hence the arcs of matter created by them are called tidal tails. 
 
The disks of spiral galaxies are, due to their extension, particularly sensitive to tidal forces. The tidal tails thereby formed mostly consist of matter coming from the disks of the galaxies, i.e. stars and a considerable amount of gas. 

Numerical calculations show that some of the gas in the tidal tails collapses into structures that are bound by their own gravity \citep{Barnes1992a,Elmegreen1993,Bournaud2010}. These structures have masses of up to $10^9 \ {\rm M}_{\odot}$ \citep{Elmegreen1993,Bournaud2006} and have radii of the order of 1 kpc \citep{Wetzstein2007,Bournaud2008}. It has also been shown that these objects can survive on a time-scale of $10^9$ years and can be sites of long-lasting star formation \citep{Bournaud2006,Recchi2007}. With these properties, the structures emerging in the tidal tails can be considered galaxies (cf. \citealt{Bournaud2007,Forbes2011}). Due to their size and their origin, such galaxies been named tidal dwarf galaxies (TDGs), and a number of structures that are TDG-candidates have been observed \citep{Mirabel1992,Monreal2007,Yoshida2008,Duc2011}.

However, the formation of the first galaxies in the Universe is driven by non-baryonic cold dark matter (CDM) according to the $\Lambda$CDM-model, which is the currently prevailing cosmological model. The CDM is thought to collapse into haloes and thereby to create the gravitational potentials that bind the baryons of the forming galaxies.  In order to distinguish them from TDGs, these galaxies are called primordial galaxies.

In contrast to the primordial galaxies, the TDGs are predicted to consist only of baryonic matter, even if the progenitors of the TDGs contained a substantial amount of CDM (\citealt{Barnes1992a,Duc2004,Bournaud2006}; see also \citealt{Bournaud2010} for a theoretical discussion of this finding).

Since only the baryonic matter interacts electromagnetically, CDM and baryonic matter must behave differently. Due to the different composition of the TDGs and the primordial galaxies, it would be natural if these two types of galaxies would constitute populations that are distinguishable by their properties. Thus, observations and theoretical calculations support the notion that there are two types of galaxies, namely primordial galaxies and TDGs. This finding has been termed the 'Dual Dwarf Galaxy Theorem' by \citet{Kroupa2012}.

A substantial fraction of the galaxies of the Universe are dwarf elliptical galaxies (dEs). These dEs are of particular interest, because the masses of their stellar populations and their radii would fit to TDGs, but they are usually considered to be the kind of galaxies that forms within CDM-haloes of rather low masses (see, e.g., \citealt{Li2010,Guo2011}). Reviews on dEs and how they may have formed are given by \citet{Ferguson1994} and \citet{Lisker2009}.

Using a compilation of data on old, dynamically hot stellar systems by \citet{Misgeld2011} on the one hand and a first-time compilation of data from various authors on masses and radii of observed TDG-candidates (observed or from numerical calculations) on the other hand, it is discussed in this paper whether primordial galaxies and TDGs are indeed distinguishable populations, as would be expected. The data used for this comparison is described in Section~(\ref{sec:Data}). The results are presented and discussed in Sections~(\ref{sec:Results}) and~(\ref{sec:discussion}). Our conclusions are given in Section~(\ref{sec:Conclusion}).

\section{Data}
\label{sec:Data}

\subsection{Old stellar systems}
\label{sec:DataOld}

\subsubsection{Galaxies}
\label{sec:galaxies}

Data on the masses and the radii of old elliptical galaxies are taken from \citet{Bender1992} and \citet{Bender1993}, \citet{Ferrarese2006}, \citet{Misgeld2008} and \citet{Misgeld2009}. The data on the dwarf spheroidal galaxies (dSphs) are taken from Table~(1) in \citet{Misgeld2011}, provided an estimate of the mass of their stellar populations, $M_{*}$, is available there. This table also lists some compact elliptical galaxies, which are included in the present compilation as well. The catalogues of galaxies in the Hydra I cluster \citep{Misgeld2008} and galaxies in the Centaurus cluster \citep{Misgeld2009} comprise a large number of dwarf elliptical galaxies (dEs) and are of particular interest for filling a gap in luminosity between the data from \citet{Bender1992,Bender1993} and the dSphs from Table~(1) in \citet{Misgeld2011}. The $M_{*}$ of all mentioned galaxies are calculated from their published luminosities and colours, using the estimates for their stellar mass-to-light ratios published in \citet{Misgeld2011}. Note that the baryonic masses, $M$, of these objects are essentially equal to $M_{*}$ since these kinds of galaxies contain almost no gas or dust \citep{Wiklind1995,Young2011}.

\subsubsection{GCs and UCDs}
\label{sec:GCsandUCDs}

Masses and effective radii of globular clusters (GCs) and ultra compact dwarf galaxies (UCDs) are taken from Table~(5) in \citet{Mieske2008}. Note that the masses listed in that table are mass estimates based on the internal dynamics of the GCs and UCDs (i.e. dynamical masses, $M_{\rm dyn}$) instead of masses estimated from the light and colour of the stellar populations (i.e. $M_{*}$). The internal dynamics of GCs and UCDs is however probably not influenced by a hypothetical presence of DM in them, since DM would usually be distributed over larger scales. \citep{Murray2009,Willman2012}. A non-Newtonian law of gravity in the limit of weak fields (i.e. the alternative to the dark matter hypothesis) would also leave the dynamics of GCs and UCDs unaffected in most cases (see figure~7 in \citealt{Kroupa2010}). Finally, GCs are free of gas and dust  \citep{vanLoon2006}, and given the similarities of UCDs to GCs, it is reasonable to assume the same for UCDs. These reasons imply that $M_{\rm dyn}=M_{*}$ for GCs and UCDs\footnote{Note that some authors discuss the elevated $M/L_V$ ratios of UCDs (see, e.g., \citealt{Hasegan2005,Dabringhausen2008,Mieske2008}), which suggest the opposite to be true. The detected difference between the $M_{\rm dyn}$ and the $M_{*}$ of UCDs is however rather small, even though this deviation probably carries important information on star formation in UCDs (see \citealt{Dabringhausen2009,Dabringhausen2012,Marks2012b} and the review by \citealt{Kroupa2011}).}.

\subsection{TDGs}
\label{sec:TDGs}

\subsubsection{Observed TDG-candidates}
\label{sec:obsTDG}

Data on the masses and the effective radii of observed TDG-candidates are difficult to obtain and collected from various sources in the literature:

\begin{itemize}
\item \citet{Miralles2012}. The sample from \citet{Miralles2012} is an extension of the sample from \citet{Monreal2007}. The TDG-candidates taken from \citet{Miralles2012} are, among all TDG-candidates considered in the present paper, the ones for which the most complete information on masses and radii is available. The data given on the TDG-candidates comprise their equivalent total radii ($r$), their effective radii ($r_{\rm e}$), their mass estimated from their $I$-band luminosity using the ages estimated under the assumption of a single star burst ($M_I$), their mass estimated from their H$\alpha$-emission lines ($M_{{\rm H}\alpha}$), their mass estimated from their $I$-band luminosity under the assumption that most stars in the TDG-candidates are old ($M_{\rm old}$) and their mass estimated from the internal dynamics of the TDG-candidates ($M_{\rm dyn}$). $M_I$, $M_{{\rm H}\alpha}$ and $M_{\rm old}$ are all estimates for $M_{*}$, the mass of the stellar population of the TDG-candidate. In oder to have a concrete value for $M_{*}$, $M_{*}=M_I$ is assumed, since estimates for $M_{*}$ based on optical luminosities are available also for all other TDG-candidates, in contrast to estimates based on H$\alpha$-emission. Setting $M_{*}=M_I$ therefore adds to the homogeneity of the sample of TDG-candidates. The adopted single-burst age is the average of the age estimate derived from photometric data and the age estimate derived from the equivalent width of the H$\alpha$-emission. The values are of the order of $10^6$ years. \citet{Miralles2012} estimate $r$ by adding up the areas of all H$\alpha$-emitting regions within a TDG-candidate, leading to a total area $A_{\rm T}$, from which $r$ is calculated from
\begin{equation}
r=\sqrt{\frac{A_{\rm T}}{\pi}}.
\label{eq:equradius}
\end{equation}
The average ratio between $r$ and $r_{\rm e}$ of the TDG-candidates in \citet{Miralles2012} is 3.5. The standard deviation about this value is 1.1. Simple estimates of $r_{\rm e}$ from $r$ can thereby be calculated from
\begin{equation}
r=(3.5 \pm 1.1)\times r_{\rm e}
\label{eq:re-estimator}
\end{equation}
for other galaxies. Note, however, that \citet{Miralles2012} only estimate the effective radius of the dominating knot if a TDG-candidate has more than one star-forming knot. Thus, $r_{\rm e}$ is underestimated for these TDG-candidates.

\item \citet{Galianni2010}. The TDG-candidates discussed by them have $V$-band luminosities of $1.6 \times 10^6 \ {\rm L}_{\odot}$ and  $2.6 \times 10^6 \ {\rm L}_{\odot}$, respectively. By giving estimates for the stellar masses of these TDG-candidates, \citet{Galianni2010} implicitly state that the $V$-band mass-to-light ratio of the TDG-candidates is $2.5 \ {\rm M}_{\odot} / {\rm L}_{\odot}$,  and $2.3 \ {\rm M}_{\odot} / {\rm L}_{\odot}$, respectively. A comparison with single-burst stellar population models (e.g. \citealt{Maraston2005}) suggests that these assumptions on the $M/L_V$-ratios of the TDG-candidates discussed in \citet{Galianni2010} are reasonable, since \citet{Galianni2010} conclude from a spectroscopical analysis that the stellar populations of their TDG-candidates are old and have metallicities [Fe/H]$>-1$, like the ones of old GCs and dSphs. Values for $r_{\rm e}$ have been found by fitting S\'{e}rsic-profiles \citep{Sersic1968} to the TDG-candidates.

\item \citet{Yoshida2008}. Photometric data suggests an age of the order of $10^8$ years for the star-forming knots they observed (termed 'fireballs' by them). This motivates their assumption of $M_{*}/L_R=1 {\rm M}_{\odot} / {\rm L}_{\odot}$ for the objects when they calculate the stellar masses of the objects from their $R$-band luminosities. Values for $r_{\rm e}$ have been estimated by fitting Gaussian profiles to their luminosity profiles. They are only given collectively as ranging between 200 pc and 300 pc for all observed objects. In order to have a concrete value for the radii, they are set to 250 pc for all fireballs in the present paper.

\item \citet{Duc2007}. The TDG-candidate identified by them has a stellar population with a mass between $M_{*}=3\times 10^7 \ {\rm M}_{\odot}$ and $M_{*}=7\times 10^7 \ {\rm M}_{\odot}$, as they find from fitting a modeled stellar population to the spectral energy distribution of the TDG-candidate. In order to have a definite value, $M_{*}=5\times 10^7 {\rm M}_{\odot}$ is assumed in the present paper. Photometric data suggests that most stars in this TDG-candidate formed $3\times 10^8$ years ago and the diameter of the TDG-candidate is given as 4200 pc.

\item \citet{Bournaud2007}. The stellar masses of the three TDGs discussed in that paper have been estimated from their optical luminosities and models of young stellar populations, since \citet{Boquien2007} estimate ages of less than $5\times 10^6$ years for the stellar populations of these galaxies. The radii given in \citet{Bournaud2007} are the radii up to which rotation curves have been measured. For an estimate of $r_{\rm e}$ from these radii, equation~(\ref{eq:re-estimator}) is used in the present paper.

\item \citet{Tran2003}. Photometric data suggest an age of $4-5\times 10^6$ for the TDG-candidate discussed in that paper. Its $V$-band luminosity (corrected for emission lines) then implies $M_{*}=6.6\times 10^5 \ {\rm M}_{\odot}$. $r_{\rm e}$ was estimated by fitting a King model \citep{King1962} to the surface-brightness profile of the object. 

\item \citet{Hunsberger1996}. The masses of the TDG-candidates listed in that paper are estimated from their $R$-band luminosities under the assumption that the $M_{*}/L_R$ ratios of the TDG-candidates is 1 ${\rm M}_{\odot} / {\rm L}_{\odot}$. This $M_{*}/L_R$ ratio implies an age of the order of $10^8$ years for the TDG-candidates. This choice for the age is motivated with \citet{Hunsberger1996} searching for TDG-candidates in Hickson compact groups \citep{Hickson1982}, i.e. in very compact groups of galaxies. Such groups have lifetimes of the order of $10^8$ years, within which the formation of TDGs is triggered by the interaction between primordial galaxies belonging to the group. The extension of the TDG-candidates is quantified in \citet{Hunsberger1996} by estimates of their diameters from their projected areas. Estimates of the $r_{\rm e}$ of the TDG-candidates are calculated in the present paper with equation eq.~(\ref{eq:re-estimator}).
\end{itemize}

The adopted properties of the observed TDG-candidates are summarized in Table~(\ref{tab:TDG}) in Appendix~(\ref{Appendix1}).

Note that not all objects in Table~(\ref{tab:TDG}) are confirmed TDGs. The reasons are the following:

\begin{itemize}
\item For some young objects, it is doubtful whether they will be stable (cf. \citealt{Monreal2007,Miralles2012}), even though their origin from tidal interactions between primordial galaxies is not disputed.

\item \citet{Tran2003} argue that the TDG-candidate they observed possibly was a stellar supercluster (SSC), i.e. a gravitationally bound complex of star clusters, which can evolve into a galaxy if a galaxy is defined as a stellar system with a relaxation time larger than a Hubble time \citep{Kroupa1998,Forbes2011}. In this sense, SSCs can be understood as precursors of TDGs. SSCs are however also seen as likely progenitors of extended star-clusters and UCDs \citep{Fellhauer2002,Fellhauer2002b,Bruens2011}, i.e. objects that are much more compact than the TDG-candidates in Table~(\ref{tab:TDG}) are, including the TDG-candidate discussed by \citet{Tran2003}.

\item \citet{Yoshida2008} consider it more likely that the objects they observed formed from gas that was stripped from the probable merger remnant RB~199 due to its motion through the intergalactic medium, rather than from matter ejected by the tidal forces acting between the progenitors of RB~199 during the merger. In order to distinguish the objects they observed from actual TDGs, they termed them 'fireballs'. On the other hand, the fireballs are gas-rich and star-forming, like the TDG-candidates observed by \citet{Monreal2007} and \citet{Miralles2012}. The fireballs are also indistinguishable from the TDG-candidates based on their masses and radii, and both kinds of objects have formed from matter that was previously bound to other galaxies, in contrast to primordial galaxies. Moreover, the arguments by \citet{Bournaud2010} for why TDGs do not contain DM would also hold for galaxies that form from stripped gas. The fireballs are therefore in the following also considered as TDGs, even if the fireballs are not actual TDGs.
\end{itemize}

In the present paper, we will concentrate on the question how the TDG-candidates will evolve if they are indeed long-lived, self-gravitating structures, as at least some (if not all) of them are. For simplicity, all objects in Table~(\ref{tab:TDG}) will thus be treated like actual, long-lived TDGs in the following. This can be motivated by the finding that they indeed make the impression of a homogeneous sample in Fig.~(\ref{fig:MassRadius17}).

\subsubsection{Numerical calculations of TDGs}
\label{sec:numTDG}

As a complement to the observed TDG-candidates, numerically calculated TDG-candidates are considered as well. The formation of TDGs during the Newtonian interaction between gas-rich galaxies has been studied with numerical calculations by many authors (e.g. \citealt{Barnes1992a,Elmegreen1993,Barnes1996,Bournaud2006,Wetzstein2007,Bournaud2008}). Detailed parameters of the resulting objects are however only available for a few exemplary objects, which come from the following sources:

\begin{itemize}
\item \citet{Bournaud2008}, who show in their figure~(5) five TDG-candidates as they appear at the end of their numerical calculation. Values for $M_{*}$ are given in that figure. In order to calculate estimates for the $r_{\rm e}$ of these objects, the absolute maximal extension of these objects and the maximal extension along the orthogonal axis were read off from this figure. These values were multiplied in order to obtain an estimate $A_{\rm T}$, which was used to calculate an equivalent radius from equation~(\ref{eq:equradius}). These equivalent radii were used to calculate $r_{\rm e}$ from equation~(\ref{eq:re-estimator}).

\item \citet{Wetzstein2007}, who describe the most massive TDG-candidate that formed in their numerical calculation in detail. If the progenitor galaxy of the TDG-candidate is scaled to the Milky Way, the total mass of the TDG-candidate is $M \approx 3.5 \times 10^8 \ M_{\odot}$. About 70 per cent of this mass is gas. Since there is no DM within the TDG-candidate (even though the progenitor galaxy was assumed to reside within a DM-halo), the mass of the stellar population of the TDG is 30 per cent of its total mass. Fitting S\'{e}rsic-profiles \citep{Sersic1968} to the calculated TDG-candidate, \citet{Wetzstein2007} estimated that the $r_{\rm e}$ of the stellar population is 700 pc for the adopted scaling. The same procedure holds $r_{\rm e} = 1400$ pc of the gaseous component of the TDG-candidate.

\item \citet{Barnes1992a}, who describe the most massive TDG that formed in their numerical calculation in detail. If the two progenitor galaxies in the numerical calculation are scaled to the Milky Way, the total mass of the TDG is $M \approx 4 \times 10^8 \ M_{\odot}$. \citet{Barnes1992a} note that there is no DM within the TDG, but they do not distinguish stars and gas in their numerical calculation. It is therefore assumed here that the ratio between gas and stars is the same as in the TDG calculated by \citet{Wetzstein2007}. An estimate for $r_{\rm e}$ of this TDG was calculated from the left panel of figure~(1) in \citet{Barnes1992a}. The TDG shown there has a diameter of $\approx 6.15\times 10^{-2}$ length units, corresponding to $\approx 2500$ pc if the progenitor galaxies are scaled to the Milky Way. Using equation~(\ref{eq:re-estimator}), this implies $r_{\rm e} \approx 350$ pc.
\end{itemize}

The adopted properties of the observed TDG-candidates are summarized in Table~(\ref{tab:TDGsim}).

\begin{figure*}
\centering
\includegraphics[scale=0.8]{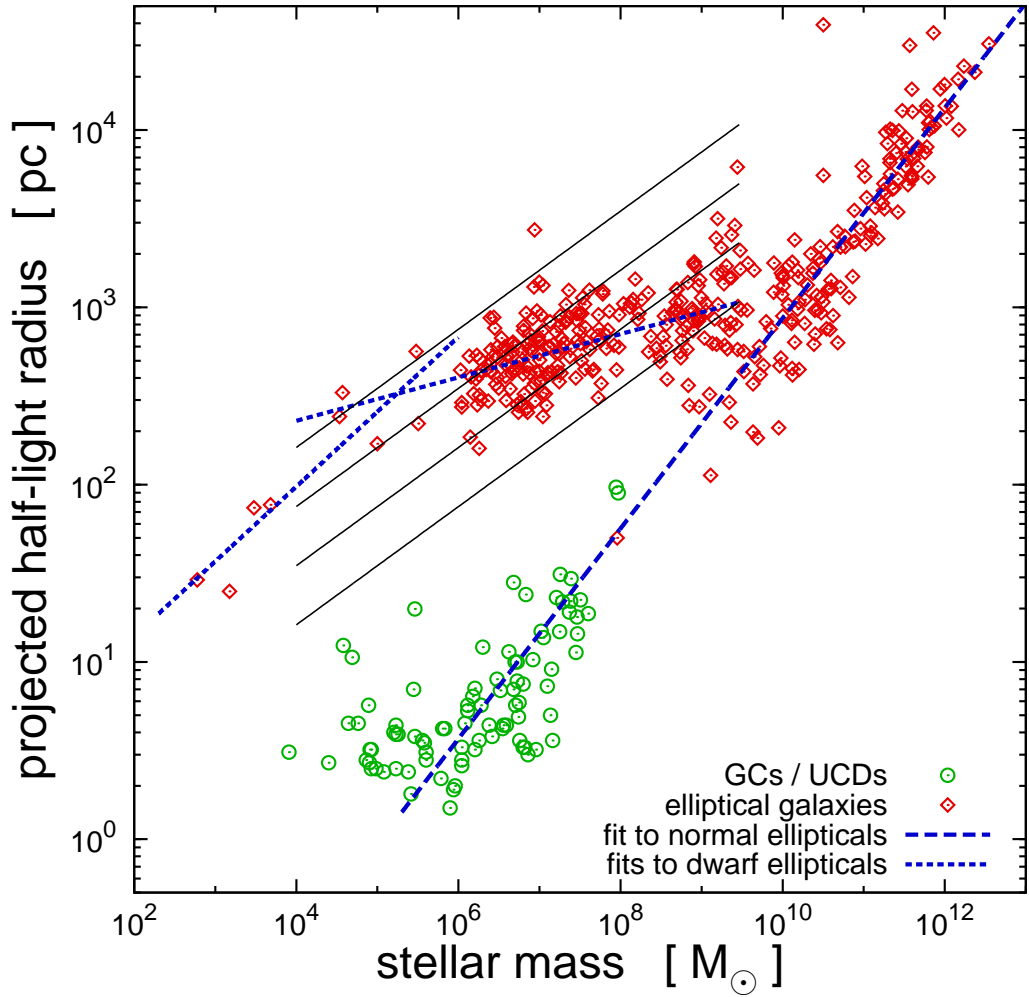}
\caption{The half-light radii of old stellar systems against the mass of their stellar populations, $M_{*}$ (cf. Section~\ref{sec:DataOld}). Provided that the mass of dust, gas and non-baryonic matter is negligible in these systems, the estimates for the mass of their stellar populations are in fact estimates of their total masses. The distinction between GCs and UCDs and elliptical galaxies is as in the literature from which the data is taken. The dashed line is a mass-radius relation obtained through a least-squares fit to (normal) elliptical galaxies with masses $M_{*}>3\times10^9 \ M_{\odot}$ (cf. equation~\ref{eq:nE}), which incidentally also fits well to the UCDs. The dotted lines are mass-radius relations obtained through least-squares fits to (dwarf) elliptical galaxies with masses $10^4 \ {\rm M}_{\odot} \le M_{*} \le 3\times10^9 \ M_{\odot}$ (cf. equation~\ref{eq:dE}), and (dwarf) elliptical galaxies with masses $M_{*}<10^6 \ {\rm M}_{\odot}$ (cf. equation~\ref{eq:dSph}), respectively. The thin solid lines indicate constant densities of $\rho=10^{-3} \ {\rm M}_{\odot} \, {\rm pc^{-3}}$, $\rho=10^{-2} \ {\rm M}_{\odot} \, {\rm pc^{-3}}$, $\rho=0.1 \ {\rm M}_{\odot} \, {\rm pc^{-3}}$ and $\rho=1 \ {\rm M}_{\odot} \, {\rm pc^{-3}}$ from top to bottom.}
\label{fig:MassRadius9}
\end{figure*}

\begin{figure*}
\centering
\includegraphics[scale=0.8]{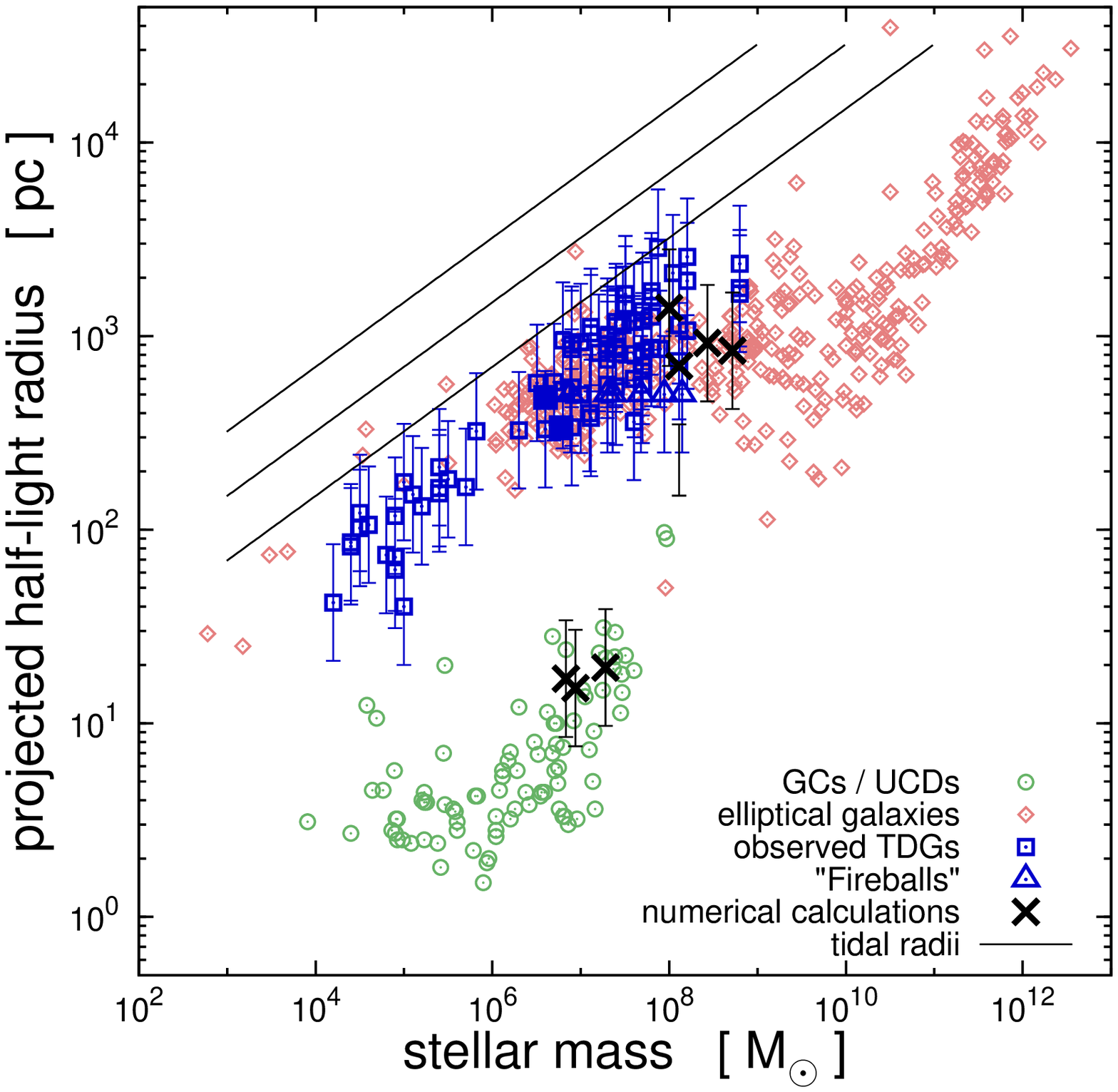}
\caption{Estimates for the final radii of TDG-candidates after their gas is expelled or has been expelled against estimates of their stellar mass, $M_{*}$. The estimates are based on the data on present-day parameters of TDG-candidates (cf. Section~\ref{sec:obsTDG} and Table~\ref{tab:TDG}) and on data on TDGs taken from numerical calculations on the formation of TDGs during the encounter of gas-rich galaxies (cf. Section~\ref{sec:numTDG} and Table~\ref{tab:TDGsim}). Probably the largest uncertainty to the data on the TDG-candidates shown here comes from the poor knowledge on how star-formation and mass-loss will influence the future evolution of the gas-rich present-day TDG-candidates until they possibly resemble old, gas-poor dEs (cf. Section~\ref{sec:evoTDG}). In order  to quantify this uncertainty on the data for the TDG-candidates, the lower limit for future radius of each TDG-candidate is taken to be its present-day radius, which corresponds to no future mass-loss according to equation~(\ref{eq:adiabatic}). The symbol representing the TDG-candidate is placed at a radius twice its present-day radius, which corresponds to a future mass-loss of half of its present-day mass according to equation~(\ref{eq:adiabatic}). The upper limit for the radius of each TDG candidate is taken to be four times its present-day radius, which corresponds to a loss of 75 per cent of its present-day mass according to equation~(\ref{eq:adiabatic}). Thus, the errorbars to the data on the TDG-candidates shown here were not formally calculated from the uncertainties to the observational  data, but represent different assumptions on the future evolution of the TDG-candidates. These assumptions are admittedly quite arbitrary, but as they are not very restrictive concerning the future mass-loss (and thus the future evolution) of the TDG-candidates, they are also conservative. The data on old stellar systems presented in Fig.~(\ref{fig:MassRadius9}) are also shown here for comparison. The thin solid lines indicate the tidal radii for stellar systems in the vicinity of a major galaxy $10^5$ pc away. The mass of the major galaxy is $M=10^{10} \ {\rm M}_{\odot}$, $M=10^{11} \ {\rm M}_{\odot}$ and $M=10^{12} \ {\rm M}_{\odot}$, from top to bottom.}
\label{fig:MassRadius17}
\end{figure*}

\section{Results}
\label{sec:Results}

\subsection{Properties of old dynamically hot stellar systems}
\label{sec:oldsystems}

It is well known that old, dynamically hot (or pressure-supported) stellar systems can be divided into two categories: A star-cluster-like population consisting of GCs and UCDs and a galaxy-like population consisting of normal elliptical galaxies (nEs), dwarf elliptical galaxies (dEs) and dwarf spheroidal galaxies (dSphs) \citep{Gilmore2007,Forbes2008,Misgeld2011}. Almost every object shown can indeed easily be assigned to one these two populations by its position in Figure~(\ref{fig:MassRadius9}). Exceptions like UCD~3 or M32 are extremely rare (see also Section~\ref{sec:UCDsTDGs}).

Within these two populations of stellar systems, subpopulations can be identified by changes of the mass-radius relations that characterize these subpopulations. This leads to a distinction between dSphs, dEs and nEs within the galaxy-like population and a distinction between GCs and UCDs within the star-cluster-like population (e.g. \citealt{Misgeld2011}).

The exact locations of the transition from one subpopulation to another is a matter of definition. In the present paper, members of the galaxy-like population are considered dEs if they have a stellar mass $M_{*} \le 3\times 10^9 \ {\rm M}_{\odot}$ and nEs otherwise. It is impossible to make a similar distinction between dEs and dSphs (cf. \citealt{Ferguson1994}). Thus dEs and dSphs will all be referred to as dEs in the following. Taking their $M_{*}$ to be equal to their total mass (cf. Section~\ref{sec:GCsandUCDs}),  members of the star-cluster-like population are considered GCs if they have a stellar mass $M_{*} \le 2\times 10^6 \ {\rm M}_{\odot}$ and UCDs otherwise.

Performing a least-squares fit to dEs with masses $M_{*}>10^4 \ {\rm M}_{\odot}$, a mass-radius relation for them is quantified as
\begin{equation}
\log_{10}\left(\frac{r_{\rm h}}{{\rm pc}}\right)=(0.122\pm0.013)\log_{10}\left(\frac{M_{*}}{{\rm M}_{\odot}}\right)+(1.87\pm0.10).
\label{eq:dE}
\end{equation}
Performing the same kind of fit to dEs with masses $M_{*}<10^6 \ {\rm M}_{\odot}$ leads to
\begin{equation}
\log_{10}\left(\frac{r_{\rm h}}{{\rm pc}}\right)=(0.42\pm0.08)\log_{10}\left(\frac{M_{*}}{{\rm M}_{\odot}}\right)+(0.3\pm0.3).
\label{eq:dSph}
\end{equation}
A constant mean density for galaxies implies
\begin{equation}
\log_{10}\left(\frac{r_{\rm h}}{{\rm pc}}\right) = \frac{1}{3}\log_{10}\left(\frac{M_{*}}{{\rm M}_{\odot}}\right)+c,
\label{eq:density}
\end{equation}
where $c$ is a constant. Equation~(\ref{eq:density}) is consistent with equation~(\ref{eq:dSph}), so that dEs with very low masses may indeed be characterized by a typical average density. This is however not the case for the more massive dEs. Their typical densities increase with their mass, as is apparent from equation~(\ref{eq:dE}).

There are however several potential problems with the data on dEs with masses $M_{*}< 10^6 \ {\rm M}_{\odot}$:

\begin{enumerate}

\item The data is very sparse in that mass range. The mass-radius relation given by equation~(\ref{eq:dSph}) is derived from only nine dEs, of which the five most massive ones are also well consistent with mass-radius relation given by equation~(\ref{eq:dE}).

\item The objects in this mass-range have very low surface brightnesses, so that their $r_{\rm e}$ are difficult to measure.

\item If these objects are pure stellar populations (for instance because they formed as tidal dwarfs and thus contain no DM), they are the most vulnerable to tidal fields that may alter their structure (\citealt{Kroupa1997,Metz2007,Casas2012}, see also Section~\ref{sec:tidal}) 
\end{enumerate}

Thus, the apparent steepening of the mass-radius relation for dEs towards lower masses has to be taken with caution.

A least-squares fit to nEs yields
\begin{equation}
\log_{10}\left(\frac{r_{\rm h}}{{\rm pc}}\right)=(0.593\pm0.027)\log_{10}\left(\frac{M_{*}}{{\rm M}_{\odot}}\right)-(2.99\pm0.30).
\label{eq:nE}
\end{equation}
This mass-radius relation for nEs is consistent with the result in \citet{Dabringhausen2008}, even though \citet{Dabringhausen2008} estimated the masses of the nEs from their $r_{\rm h}$ and their velocity dispersions, whereas in this paper their luminosities and colours were used.

UCDs lie along the same mass-radius relation as nEs, even though UCDs belong to the star-cluster-like population while nEs belong to the galaxy-like population \citep{Dabringhausen2008}. Note that UCDs were not included in the fit of equation~(\ref{eq:nE}) to the data. The mass-radius relation of GCs, in contrast, is essentially flat.

Possible reasons for the transition from dEs to nEs are discussed in Section~(\ref{sec:mrdEs}) and possible reasons for the transition from GCs to UCDs are discussed in Section~(\ref{sec:mrGCs}).

\subsection{Properties and evolution of TDGs}
\label{sec:evoTDG}

A comparison of the numerically calculated TDG-candidates with the observed TDG-candidates shows that the estimates of $M_{*}$ and $r_{\rm e}$ of the calculated TDG-candidates are consistent with the observed ones. If these parameters are however compared to the according present-day parameters of the GCs, UCDs, dEs and nEs (i.e. the stellar systems introduced in Section~\ref{sec:DataOld} and shown in Figure~\ref{fig:MassRadius9}), the (young) TDG-candidates are on a mass-radius relation below the mass-radius relation for (old) dEs. The old TDGs by \citet{Galianni2010} are however consistent with being typical dEs.

In order to find the actual interrelations between the TDG-candidates and the other stellar systems, it is necessary to estimate what they would look like if they all had the same age. This requires to account for the future evolution of the TDG-candidates listed in Tables~(\ref{tab:TDG}) and~(\ref{tab:TDGsim}) in the Appendix, since almost all of them have ages of the order of $10^8$ years or less, while the objects shown in Figure~(\ref{fig:MassRadius9}) are at least a few $10^9$ years old (see \citealt{Misgeld2011}). The age difference between these systems is consistent with the finding that many TDG-candidates show evidence for ongoing star formation, as the H$\alpha$-emission from these systems indicates.

The amount of gas is actually very substantial in the three TDG-candidates observed by \citet{Bournaud2007}. They estimate the mass of the stars in these TDG-candidates by their luminosity and the amount of gas in them by the strength of their emission lines. They thereby find that stars make up only about ten per cent of the baryonic mass in each of the TDG-candidates they studied. Such a detailed analysis of their composition has not been made for the other TDG-candidates discussed in the present paper. It is however likely that the composition of other very young TDG-candidates from \citet{Miralles2012} with ages of the order of $10^6$ years is similar, especially since the masses derived from their internal dynamics are much higher than the masses derived from the luminosity of their stellar populations. The TDG-candidates discussed by \citet{Yoshida2008} and \citet{Duc2011} suggest that also TDG-candidates with ages of at least $10^8$ years can still contain enough gas to be sites of very recent star-formation. Thus, gas is the principal mass component in many of the TDG-candidates, since they do not contain DM \citep{Barnes1992a,Bournaud2010,Kroupa2012}. The fate of the gas in the TDG-candidates is therefore decisive for their future evolution.

The two TDG-candidates studied by \citet{Galianni2010} are the only ones in Table~(\ref{tab:TDG}) that have ages similar to those of the old stellar systems introduced in Section~(\ref{sec:DataOld}). Like the other old stellar systems, the TDG-candidates show no traces of ongoing star-formation, which suggests that there is no gas left in them. This motivates the assumption that the other TDG-candidates will also have lost their gas once they have reached that age.

There are several processes by which the gas can disappear from the young TDG-candidates on a time-scale of the order of $10^9$ years:

\begin{enumerate}
\item The conversion of gas into stars.

\item The removal of the gas through heating by massive stars that form in the TDG-candidates.

\item The removal of the gas through ram-pressure stripping when the TDG-candidates move through the intergalactic medium.
\end{enumerate}

More than one of these processes may contribute to the removal of the gas. The first two processes are even intimately linked to each other, since the presence of massive stars implies recent star formation, while the removal of gas through the heating of these massive stars influences further star formation. This interrelation is known as feedback. The second and the third process are similar regarding their effect on the further evolution of the stellar system. Both imply that the TDG-candidate looses mass and in consequence expands (see \citealt{Kroupa2008} for a detailed discussion on the effect of mass-loss on stellar systems). However, the actual contribution of each of these processes to the disappearance of the gas in the TDGs is unknown. It is likely to be different for each individual TDG, but according to the models by \citet{Recchi2007}, an isolated TDG self-regulates its star-formation such that it is relatively stable against feedback and major blow-outs do not arise.

Let us assume for now that the second and the third process are the most relevant ones for the future evolution of an existing young TDG-candidate, i.e. that it looses much of its mass through the removal of gas, while only little of the gas is converted into stars that add to the existing stellar population of the TDG-candidate. This assumption is motivated by the finding that the gravitational potentials of the TDG-candidates are rather shallow \citep{Bournaud2010}, so that matter can quite easily be removed from them.

If the evolution of a stellar system is primarily driven by mass-loss, it is decisive whether the mass-loss is fast or slow compared to the crossing time of the stellar system. This crossing time can be defined as
\begin{equation}
t_{\rm cr}=\frac{2 r_{\rm e}}{\sigma},
\label{eq:crossing}
\end{equation}
where $r_{\rm e}$ is the effective radius of the stellar system and $\sigma$ is its internal velocity dispersion \citep{Kroupa2008}. For the TDG-candidates, typical values for $\sigma$ are of the order of 10 km/s (see table~2 in \citealt{Miralles2012}) and typical values for $r_e$ the order of 100 pc (see Table{\ref{tab:TDG}}). Noting that 1~km/s is essentially equal to 1~pc/$10^6$years, the typical crossing times of TDG-candidates are of the order of $10^7$ years. The TDG-candidates observed by \citet{Yoshida2008} are of the order of $10^8$ years old, but still contain enough gas for star formation. Taking this as evidence that the time scale on which the gas is lost from the TDG-candidates is not less then $10^8$ years, the mass-loss would be slow compared to the typical crossing times of the TDG-candidates. In the case of slow (adiabatic) mass loss, the expansion of a stellar system is given by
\begin{equation}
\frac{M_{\rm f}}{M_{\rm i}}=\frac{r_{\rm i}}{r_{f}},
\label{eq:adiabatic}
\end{equation}
where $M_{\rm f}$ is the final mass of the stellar system, $M_{\rm i}$ is the initial mass of the stellar system, $r_{\rm f}$ is the final radius of the stellar system and $r_{\rm i}$ is the initial radius of the stellar system \citep{Kroupa2008}.

In order to calculate the future expansion of the young TDG-candidates with equation~(\ref{eq:adiabatic}), $M_{\rm f}=M_{*}$ and $r_{\rm i}=r_{\rm e}$ is assumed. Two cases are considered here for $M_{\rm i}$, namely $M_{\rm i} = 2 M_{\rm f}$ and $M_{\rm i} = 4 M_{\rm f}$. Thus, the TDG-candidates are assumed to loose 50 per cent and 75 per cent of their mass through the removal of their gas. That is somewhat less than the amount of gas traced by \citet{Bournaud2007} in TDG-candidates, which reflects the expectation that some of the gas available at the present will not be expelled, but will be converted into stars. The resulting final parameters of the TDGs are shown in Fig.~\ref{fig:MassRadius17}. 

Given the assumptions that were used to calculate them, these estimates for the final values for $M_{*}$ and $r_{\rm e}$ of the young TDG-candidates can only be approximations. Their overall consistency with the according parameters for the old TDGs discussed by \citet{Galianni2010} and the (old) dEs is however remarkable.

Despite the fact that young TDGs probably mostly consist of gas, it is by no means clear that gas-expulsion is indeed as important as assumed for the estimation of their final parameters as shown in Figure~(\ref{fig:MassRadius17}). Gas expulsion is however not the only process by which the extension of the stellar component of a TDG can grow. According to numerical calculations performed by \citet{Recchi2007}, star formation in a TDGs starts at the centre and spreads from there with time.

Whether this buildup of the stellar population of the TDG from the inside to the outside or mass loss is more important for the evolution of its size is unclear at the present. It would however be natural that growth from the inside to the outside is most relevant for the TDGs with the deepest potentials (i.e. the most massive ones), as they are the least vulnerable to mass loss. Thus, young TDGs lie on a mass-radius sequence below the one of old dEs, but the parameters of the TDG-candidates would evolve naturally towards the parameters of dEs as they reach a comparable age.

\subsection{The tidal radii of the TDGs}
\label{sec:tidal}

A TDG expelled from the interacting or merging progenitor galaxies (cf. \citealt{Elmegreen1993}) will evolve self-regulated \citep{Recchi2007} and may become a dwarf irregular galaxy \citep{Hunter2000}. However, if a TDG is bound to a host (either a larger galaxy or a galaxy cluster), its size is limited by its tidal radius. This tidal radius, $r_{\rm tid}$, depends on the mass of the TDG, $M_{\rm gal}$, the mass of the host, $M_{\rm host}$, and the distance between the TDG and its host. For systems that effectively are point masses and obey $M_{\rm host} \apprge 10 \ M_{\rm gal} $, a good approximation to $r_{\rm tid}$ is given by \citep{Binney1987}
\begin{equation}
r_{\rm tid}=\left(\frac{M_{\rm gal}}{3M_{\rm host}}\right)^\frac{1}{3}R.
\label{eq:tidal}
\end{equation}
For given values for $R$ and $M_{\rm host}$, equation~(\ref{eq:tidal}) gives the minimum average density a TDG needs to have in order to be an object kept together by its own gravity. At radii $r>r_{\rm tid}$, matter cannot be bound exclusively to the TDG, but only to the common gravitational potential of the TDG and its host.

Since TDGs form from tidal arms, it is indeed likely that a TDG is bound to a larger structure. This larger structure  can be its progenitors or a galaxy cluster in which the TDG formed. The notion of many TDGs remaining bound to their progenitor is supported by the dwarf galaxies bound to the Milky Way, whose disk-like distribution and aligned angular momenta can, as it seems, only be understood if they are ancient TDGs \citep{Kroupa2005,Metz2008,Kroupa2010,Pawlowski2012a,Pawlowski2012b,Kroupa2012}. For TDGs that remain bound to their progenitor galaxies total masses $10^{10} \ {\rm M}_{\odot} \apprle M_{\rm host} \apprle 10^{12} \ {\rm M}_{\odot}$ (i.e. the mass of a major galaxy) and distances $R \approx 10^5$ pc would be typical. The tidal radii implied by these parameters are plotted in Fig.~\ref{fig:MassRadius17}. 

If the TDG-candidates introduced in Section~(\ref{sec:TDGs}) adiabatically loose 75 per cent of their mass (i.e. $M_{\rm i} = 4 M_{\rm f}$ in equation~\ref{eq:adiabatic}), their radii become similar to the tidal radii shown in Figure~(\ref{fig:MassRadius17}). Thus, the TDG-candidates cannot expand any further if they loose even more mass, provided that the choices of $M_{\rm host}$ and $R$ are appropriate. Note however that a adiabatically expanding TDG would not dissolve completely if its outskirts expand beyond its tidal radius.

Interestingly, the tidal radii shown in figure~(\ref{fig:MassRadius17}) also coincide well with the maximum $r_e$ observed for dEs with stellar masses $M_{*} \apprle 10^7 {\rm M}_{\odot}$. This is evidence that tidal fields are indeed relevant for the structure of low-mass galaxies for which dark matter haloes do not play a role, as is very likely to be the case for such galaxies (see \citealt{Kroupa2012} and Sections~\ref{sec:dEsTDGs} and~\ref{sec:mrdEs}). At higher masses, the mass-radius relation for dEs is significantly flatter than a relation implying a constant average density in dEs. Thus, the density of the more massive dEs tends to increase with their mass, so that they are less effected by tidal fields. The reason might be that the galaxies with the lowest masses have the weakest gravitational potentials and therefore are the most vulnerable to mass-loss and subsequent expansion. More massive galaxies might keep more of their initial gas and use it up in star formation, so that they expand less.

Thus, the young TDG-candidates are likely to expand (see Section~\ref{sec:evoTDG}), but tidal fields are likely to limit this expansion. The young TDG-candidates discussed in this paper would thereby naturally evolve onto the mass-radius relation of dEs and become indistinguishable from them in this respect.

\section{Discussion}
\label{sec:discussion}

\subsection{The relation between dEs and TDGs}
\label{sec:dEsTDGs}

According to the $\Lambda$CDM-model, there are two kinds of galaxies with masses $10^{6} \ {\rm M}_{\odot} \apprle M \apprle 10^{10} \ {\rm M}_{\odot}$. The first kind are primordial dwarf galaxies that form within DM-haloes of rather low masses \citep{Li2010,Guo2011}, as they are predicted in numerical calculations of structure formation in the Universe. The second kind are TDGs, whose formation is predicted in numerical calculations of encountering galaxies that are set up in concordance with the $\Lambda$CDM-model \citep{Barnes1992a,Bournaud2006}. This prediction by the $\Lambda$CDM-model has been termed the 'Dual Dwarf Galaxy Theorem' by \citet{Kroupa2012}.

The 'Dual Dwarf Galaxy Theorem' poses a problem for the $\Lambda$CDM-model for several reasons:

\begin{enumerate}
\item It would be natural that primordial galaxies containing a substantial amount of CDM have a different structure than old TDGs, which do not contain DM and are of a different origin. Thus, old TDGs and primordial galaxies would be expected to form two distinct populations. Following this argument, the data in Fig.~(\ref{fig:MassRadius17}) suggests that the dEs are old TDGs. The dynamical $M/L$-ratios in the central parts of dEs with masses $10^8 \ {\rm M}_{\odot} \apprle M_{*} \apprle 10^9 {\rm M}_{\odot}$ imply that there is little CDM at best in these regions of the dEs \citep{Wolf2010,Toloba2011,Forbes2011b}. Admittedly, about a (hypothetical) presence of CDM in the outskirts of these galaxies, nothing is known so far from observations. Less massive dEs tend to have seemingly higher $M/L$-ratios, but this may be due to the disturbance from a tidal field of a neighbouring major galaxy \citep{Kroupa1997,Casas2012}, or due to Newtonian gravity not being valid in the limit of very weak gravitational fields (see figure~8 in \citealt{Kroupa2010}). UCDs are galaxies according to some definitions of a galaxy (see \citealt{Forbes2011} and Section~\ref{sec:UCDsTDGs}) and have elevated mass-to-light ratios, but a significant amount of CDM in them is very unlikely nevertheless (\citealt{Murray2009,Willman2012}; see also \citealt{Dabringhausen2012}). Thus, a population of dwarf galaxies that definitively formed within DM-haloes cannot be identified in Fig~\ref{fig:MassRadius17}. 

\item Even if all dEs were galaxies that formed within low-mass CDM-haloes, their number would still be low compared to the predicted number of CDM-haloes in the appropriate mass-range \citep{Moore1999,Klypin1999}; a finding that has been termed the 'missing satellite problem'. In consequence, mechanisms that would supress the formation of galaxies within most low-mass CDM-haloes were discussed (e.g. \citealt{Benson2002,Li2010}). The number of dwarf galaxies that form nevertheless according to such models is however only consistent with the number of observed dwarf galaxies if the existence of TDGs is neglected \citep{Kroupa2010,Kroupa2012}. There is however strong observational evidence for formation of TDGs in encounters between galaxies (e.g \citealt{Mirabel1992}) and that the TDGs thereby created are numerous (e.g. \citealt{Hunsberger1996}). Thus, the 'Missing Satellite Problem' persists.

\item The satellite galaxies of the Milky Way form a rotationally (or angular-momentum) supported disk \citep{Kroupa2005,Metz2008,Pawlowski2012a}, which would be logical if these galaxies are TDGs, but incomprehensible if they formed as primordial structures in agreement with the $\Lambda$CDM-model \citep{Pawlowski2012b}. This implies that all dEs around the Milky Way are in fact ancient TDGs. This finding strengthens the previous two points, namely that firstly all dEs are more likely old TDGs rather than primordial galaxies and that secondly the 'Missing Satellites Problem' thereby is far from being solved within the standard cosmological model.
\end{enumerate}

The notion that all dEs are old TDGs raises the question whether a sufficiently high number could have been produced over the age of the Universe. Concerning this matter,  \citet{Bournaud2006} show in numerical calculations that about 25 per cent of the TDGs initially created in an encounter would survive for more than $2\times10^9$ years, which corresponds to an average between 1 and 2 long-lived TDGs per calculated interaction. \citet{Okazaki2000} argue that a TDG-production at this rate would already be sufficient to account for all dwarf galaxies in the nearby Universe (also see \citealt{Kroupa2010}).

An implication of all dEs being ancient TDGs is that the dynamical $M/L$-ratios of dEs with masses $M_{*}\apprge10^8 \ {\rm M}_{\odot}$ must be consistent with the $M/L$-ratios of pure stellar populations. This can be seen in figure~7 in \citet{Misgeld2011}, which shows that the internal gravitational acceleration in these dEs is at or above the limit for Milgromian dynamics (\citealt{Milgrom1983}, see \citealt{Famaey2011} for a rewiev), while TDGs do not contain dark matter even if their progenitors did \citep{Barnes1992a,Bournaud2010}. 

Observations of dEs with dynamical masses $M_{\rm dyn} \apprge 10^8 \ {\rm M}_{\odot}$ reveal that most of them have dynamical $I$-band $M/L$-ratios of $2 \apprle M_{\rm dyn}/L_I \apprle 4$ within their effective radii \citep{Wolf2010}. Such $M_{\rm dyn}/L_I$-ratios correspond, for instance, to the $M_{*}/L_I$ ratios of single-burst stellar populations with a metallicity of [Z/H]$=-0.33$ and ages $5 \times 10^9 \ {\rm yr} < t < 13 \times 10^9 \ {\rm yr}$ \citep{Maraston2005}. This does not exclude that the actual stellar populations of the dEs shown in \citet{Wolf2010} are more luminous, so that an additional matter component would be needed in order to explain the observed $M_{\rm dyn}/L_I$ ratios. The central $M_{\rm dyn}/L_I$-ratios of the 21 dEs listed in table~7 in \citet{Toloba2011} are however indeed of the same order of magnitude as their central stellar $M_*/L_I$-ratios. Given that the central $M_{*}/L_I$-ratios of these dEs are quite uncertain and in 7 cases higher than the according $M_{\rm dyn}/L_I$-ratios, there is moreover no compelling evidence that $M_{\rm dyn}>M_{*}$ holds for them. Thus, observations do indeed support the notion that the dynamics of the central region of dEs is consistent with Newtonian dynamics, even if little or no dark matter is present there.

The typical $M_{\rm dyn}/L$-ratios of dEs with masses $M_{*} \apprle 10^8$ (often referred to as dSphs in the literature, but see \citealt{Ferguson1994} and Section~\ref{sec:oldsystems}) strongly increase with decreasing mass (cf. figure~4 in \citealt{Wolf2010}). This makes them inconsistent with the assumption that dEs with masses $M_{*} \apprle 10^8$ are pure stellar populations that are in virial equilibrium and obey to Newtonian dynamics. However,  being less tightly bound than the more massive dEs (cf. figure~7 in \citealt{Misgeld2011}), these dEs are more likely to be disturbed by tidal fields, which would lead to seemingly high $M_{\rm dyn}/L$-ratios if the dEs are assumed to be in virial equilibrium \citep{Kroupa1997,McGaugh2010,Casas2012}. Moreover, the internal accelerations in low-mass dEs are in the regime where Milgromian dynamics \citep{Milgrom1983} would be relevant (cf. figure~7 in \citealt{Misgeld2011}). Note that Milgromian dynamics would also explain the remarkably high dynamical masses of the TDG-candidates discussed by \citet{Bournaud2007}, which cannot be explained with the baryonic matter found in these galaxies, even though numerical experiments strongly predict the absence of DM in TDGs (e.g. \citealt{Barnes1992a,Gentile2007,Bournaud2010}). Thus, also low-mass dEs can be understood as dark-matter free TDGs, if they are not in virial equilibrium or if their internal dynamics is non-Newtonian.

It has also been established from observations that less luminous elliptical galaxies (i.e. dEs) tend to be bluer than more luminous elliptical galaxies (i.e. nEs). This colour-magnitude relation exists, because less luminous elliptical galaxies tend to be younger and less metal-rich than more massive elliptical galaxies \citep{Gallazzi2006}.

If the notion of all dEs being old TDGs is correct, and nEs are primordial galaxies, the dEs would tend to be younger than the nEs because the dEs could as a matter of principle only form after the formation of the first primordial galaxies. Thus, this scenario would naturally explain 'downsizing', i.e. that the least massive galaxies tend to have the youngest stellar populations, although they should be the oldest galaxies according to the $\Lambda$CDM model (see, e.g., \citealt{Cimatti2004}).

Understanding why the dEs tend to be less metal-rich than nEs is less intuitive under the premise that dEs are old TDGs. This is because the dEs would have formed from pre-enriched material if they are not primordial objects. However, the old TDGs have formed at a time when the primordial galaxies were less self-enriched than they are today. Old low-mass and metal-poor dEs can therefore be understood as TDGs that formed from matter that was scarcely pre-enriched and in which self-enrichment was not very effective. The low-mass dEs are indeed less tightly bound than the more massive dEs and nEs, as figure~7 in \citet{Misgeld2011} indicates. Thus, the low-mass dEs are may have been more likely to loose the gas expelled by evolving stars,  while high-mass dEs and nEs may have been more likely to reprocess it. This would explain why the typical metallicity of dEs increases with their mass, no matter whether they are ancient TDGs or not.

Thus, so far the properties of the dEs seem to be consistent with them being old TDGs. This would, however, imply that there are no primordial galaxies with stellar masses $M_{*} \apprle 10^{10} \ {\rm M}_{\odot}$, which would be inconsistent with the $\Lambda$CDM-model \citep{Kroupa2010,Kroupa2012}. This may be evidence for the $\Lambda$CDM-model needing to be replaced with a cosmological model where the apparent need for DM as an explanation for the internal dynamics of galaxies is replaced with a non-Newtonian law of gravity in the ultra-weak field limit. An excellent example of such a law of gravity is provided by Milgromian dynamics \citep{Milgrom1983}.

\subsection{The relation between UCDs and TDGs}
\label{sec:UCDsTDGs}

The highly resolved numerical calculation of the merger of two galaxies performed by \citet{Bournaud2008} implies that two types of stellar systems are created during the merger:

\begin{enumerate}
\item pressure-supported stellar systems with masses $10^5 \ {\rm M}_{\odot} < M_{*} < 10^7 \ {\rm M}_{\odot}$ and diameters between 10 and 100 pc. \citet{Bournaud2008} identifies this type of stellar system with the super star clusters (SSCs) discussed by \citet{Kroupa1998}, i.e. complexes of star clusters that are kept together by mutual gravitational forces. Possible SSCs have been observed by \citet{Whitmore1995} and they will evolve into objects that observers would classify as UCDs \citep{Kroupa1998,Fellhauer2002} or extended star-clusters \citep{Fellhauer2002b,Bruens2011}.

\item rotating stellar systems with masses $10^8 \ {\rm M}_{\odot} < M_{*} < 10^9 \ {\rm M}_{\odot}$ and diameters of a few $10^3$ pc. This type of stellar system has been identified with 'classical' TDGs by \citet{Bournaud2008}. Candidates for observed TDGs are listed in Table~(\ref{tab:TDG}), and may evolve into dEs (see Sections~\ref{sec:evoTDG} and~\ref{sec:tidal}).
\end{enumerate}

The properties of the stellar systems that form during a merger of galaxies according to the numerical calculations by \citet{Bournaud2008} are therefore consistent with observations, as is also illustrated in Figure~(\ref{fig:MassRadius17}). This consistency with the observations includes the lack of objects intermediate to SSCs and 'classical' TDGs. Given the probable future evolution of these objects, their absence translates into the absence of objects intermediate to UCDs and dEs, which is illustrated with Figure~(\ref{fig:MassRadius9}).

Thus, the numerical calculation by \citet{Bournaud2008} correctly reproduces the mass spectrum and the sizes of objects forming during a galaxy merger. This is strong evidence for the physical processes included in the model (gas dynamics, stellar dynamics, star formation law) being sufficient and their implementation in the numerical code being adequate for the overall description of a galaxy merger. A detailed understanding of why two distinct types of objects are formed during the merger (namely SSCs and TDGs) is however still missing. Given the apparent link to physical processes, it nevertheless stands to reason to distinguish galaxies from star-clusters by their different structure, i.e. by the gap between a star-cluster-like population (to which the UCDs belong) and a galaxy-like population (to which the TDGs belong). This is essentially equivalent to the distinction between star-clusters and galaxies proposed by \citet{Gilmore2007}.

Note however that there are also other ways to define a galaxy (see \citealt{Forbes2011}). The choice of the definition is decisive for the classification of UCDs and their progenitors.

If a galaxy is defined as a stellar system whose dynamics cannot be explained with its baryons obeying Newton's laws of gravity \citep{Willman2012}, UCDs\footnote{The elevated $M/L_V$ ratios of UCDs \citep{Hasegan2005,Dabringhausen2008,Mieske2008} are probably not due to CDM or non-Newtonian gravity (see Section~\ref{sec:GCsandUCDs}), but to the presence of a large population of neutron stars and stellar-mass black holes in UCDs, which is the consequence of a top-heavy stellar initial mass function in UCDs which formed as a major star-burst \citep{Dabringhausen2009,Dabringhausen2012,Marks2012b}.} and SSCs are star-clusters. The internal properties of even one of the most massive UCDs have indeed been argued to be consistent with it being an extremely massive star cluster \citep{Frank2011}.

If a galaxy is defined as a stellar system whose relaxation time at its half-mass radius, $t_{\rm }$, is longer than the age of the universe, $\tau_{\rm H}$,  \citep{Kroupa1998,Kroupa2012}, most UCDs and SSCs are galaxies. Thus, the numerical calculations performed by \citet{Bournaud2008} would predict the formation of two different kinds of tidal galaxies according to this definition of a galaxy. Note however that this definition implies that any stellar system will become a star-cluster at some point of time by the aging of the universe.

In essence, each of the proposed definitions is based on a property that is typical for a galaxy. By choosing a certain definition, the importance of the according property is emphasized. Defining a galaxy as a stellar system with $t_{\rm rh} > \tau_{\rm H}$ emphasizes the fact that such systems cannot have evolved dynamically through two-body encounters at the present, which has important implications for how to model the dynamical evolution of such systems effectively \citep{Kroupa2012}. Defining a galaxy by its extension, or by the impossibility to explain its dynamics with its baryonic matter and the Newtonian laws of gravity, emphasizes a fundamental physical difference of these systems.

\subsection{The GCs of dEs}

It is known that dEs usually are surrounded by GC systems. Typical sizes of these GC systems range from a few GCs to about 100 GCs. Not considering the total number of GCs in these systems, but the number of GCs per unit luminosity of their host galaxy, the GC systems of some low-luminosity dEs are actually large in comparison to other galaxies \citep{Peng2008}.

If dEs are ancient TDGs, their GCs can have formed in different ways:
\begin{enumerate}
\item The numerical calculation by \citet{Bournaud2008} suggests that during a galaxy merger GCs and TDGs are created at the same time. If they are formed within the same phasespace volume, a forming TDG might capture forming GCs within its gravitational field.

\item SSCs, which are possible progenitors of TDGs, are highly substructured. While most of the subsystems quickly  merge into an object that would be identified as a UCD or a TDG by observers, some subsystems may survive on a timescale of $10^9$ years (see figure~11 in \citealt{Fellhauer2002c}) and might qualify as GCs.

\item According to \citet{Weidner2004}, the mass of the most massive star-clusters that can form within a stellar system depends on the star formation rate of that stellar system. If the initial star formation rate in the TDGs was high enough, GCs may have formed during the evolution of the TDG as its most massive star-clusters. In contrast to the first two scenarios, this scenario implies that the GCs are younger than its host.
\end{enumerate}

In any case, GC-candidates should be very common around TDG-candidates of any age, if the dEs are old TDGs and if their GC-systems form early during their evolution. As these GCs would have formed from pre-processed matter, they would tend to be more metal-rich than GCs that formed with the formation of a primordial galaxy. This stands in contrast with the finding that dEs tend to have a higher fraction of blue GCs than nEs, which could be interpreted as the GCs of dEs tending to be less metal-rich than the GCs of nEs (cf. figure~8 in \citealt{Peng2008}). Note however that the bluer colour of the GCs belonging to dEs could also indicate a lower age. This would be natural if the nEs are primordial galaxies while the dEs are old TDGs (and thus younger than nEs) and if the GCs formed together with their host galaxies. Moreover, metal-enrichment might not have proceeded very far in the progenitor galaxies when the progenitors of present-day dEs possibly have formed as TDGs (cf. Section~\ref{sec:dEsTDGs}). Thus, finding GC-candidates in a systematic search around TDG-candidates would be supportive evidence for the dEs being old TDGs.

If no GC-candidates are found around young TDG-candidates, this could be explained in different ways:
\begin{enumerate}
\item Only the dEs with very few or no GCs are ancient TDGs.
\item GCs form rather late during the evolution of a young TDG into a dE.
\item \citet{Marks2010} note that metal-rich GCs form with larger radii. These GCs are thus more susceptible to destruction than the GCs that formed at the age of ancient TDG-formation, which arguably formed from matter that was barely pre-enriched. Presently forming TDGs may thus have a small specific frequency of GCs.
\end{enumerate}
Note however that the case of finding no GC-candidates around TDG-candidates does not seem very likely, considering the substructure found in SSCs and the TDG-candidates discussed in \citet{Miralles2012} and the possibility that some of these substructures may survive for a long time according to \citet{Fellhauer2002c}.

\subsection{Mass-radius relations}

In the following, possible reasons for the mass-radius relations described in Section~(\ref{sec:oldsystems}) are discussed.

\subsubsection{The mass-radius relation of nEs and UCDs}
\label{sec:mrUCDsnEs}

The mass-radius relation for UCDs and nEs is very remarkable because it bridges the gap between the galaxy-like population and the star-cluster-like population (cf. Section~\ref{sec:UCDsTDGs}). The common mass-radius relation suggests that the overall structure of UCDs and nEs was shaped by a process that is relevant for both types of stellar systems, even though the formation of objects intermediate to UCDs and nEs is inhibited. As spheroidal systems with little or no substructure and primarily old stellar populations, UCDs and nEs share indeed many similarities, despite being separated by the size gap noted by \citet{Gilmore2007}. Since there is most probably no CDM in UCDs (cf. Section~\ref{sec:UCDsTDGs}), considering UCDs and nEs as similar objects at different masses argues against the presence of DM in nEs. In fact, strong evidence for the absence of CDM in galaxies has already been found by \citet{Disney2008}.

The process that shaped UCDs and nEs could be monolithic collapse, i.e. that the stellar systems form rapidly by the collapse of a single gas cloud. If the mass of a cloud is sufficient for the formation of a UCD, it becomes optically thick for infrared radiation during the collapse, which leads to internal heating. This internal heating halts the collapse and \citet{Murray2009} finds that the radius at which the collapse is halted depends on the mass of the cloud. This dependency is quantified as
\begin{equation}
\log(r_{\rm h}) \propto 0.6\log(M_{*}),
\label{eq:Murray}
\end{equation}
i.e. up to a constant by the same mass-radius dependency that was found for nEs and UCDs from their observed parameters (cf. \citealt{Dabringhausen2008,Misgeld2011} and equation~\ref{eq:nE}). Note that \citet{Murray2009} only discusses the difference between GCs and UCDs. However, if monolithic collapse is the reason for the mass-radius relation of UCDs, the fact that nEs and bulges\footnote{Note that the location of bulges within the fundamental plane suggests that they are essentially identical to nEs (e.g. \citealt{Bender1992}).} lie on the same mass-radius relation suggests that they also formed through monolithic collapse. Monolithic collapse has indeed already been considered for the formation of nEs and bulges (e.g. \citealt{Elmegreen1999,Sanders2008}).

As an extension to the model of pure monolithic collapse, it can be assumed that the gas clouds formed substructures while they collapsed. The present-day UCDs and nEs would then have formed by the merging of these substructures, as discussed in the literature for UCDs (e.g. \citealt{Kroupa1998,Fellhauer2002,Bruens2011}).

\subsubsection{The mass-radius relation of GCs}
\label{sec:mrGCs}

The members of the star-cluster-like population lie an a mass-radius relation that changes its slope at a mass $M_{*} \approx 2 \times 10^6 \ M_{\odot}$. This change of the slope marks the transition from GCs to UCDs. \citep{Hasegan2005,Mieske2008}.

The progenitors of UCDs were gas clouds above a certain mass threshold, at which gas clouds become optically thick for infrared radiation when they collapse and form stellar systems. They follow equation~(\ref{eq:Murray}). The progenitors of GCs and open star clusters like the Plejades or the Orion Nebula Cluster would be gas clouds that remained transparent for infrared radiation because their masses were below the threshold (\citealt{Murray2009}; see also Section~\ref{sec:mrUCDsnEs}). Such clouds collapse to sizes of 0.1 pc \citep{Marks2012}, which is the observed size of dense cloud cores that are thought to be the progenitors of low-mass star clusters \citep{Bergin2007}. The 0.1~pc~scale may be set by the width of filaments within molecular clouds. Star formation is observed in the filaments if the mass per unit length exceeds a critical value \citep{Andre2010}.

Thus, there is no fundamental difference between the progenitors of present-day GCs and the progenitors of present-day UCDs, except for their mass, which has implications for their evolution. A common origin of both types of stellar systems is indeed implied by the continuous transition between GCs and UCDs. This common origin may be that they have formed during the interaction between galaxies, which is not only a likely trigger for the formation of UCDs \citep{Fellhauer2002}, but also for the formation of 'classical' GCs \citep{Zepf1993}. Thus, UCDs can be understood as the most massive GCs \citep{Mieske2012}.

Such an origin would make UCDs and GCs similar to TDGs. This makes it even harder to understand why the radii of TDGs are about an order of magnitude larger than the radii of GCs and UCDs of comparable mass (see Fig~\ref{fig:MassRadius17}), so that GCs and UCDs on the one hand and TDGs on the other hand are distinct populations of stellar systems (cf. Section~\ref{sec:UCDsTDGs}). 

\subsubsection{The mass-radius relation of dEs}
\label{sec:mrdEs}

At a stellar mass $M_{*} \approx 10^{10} \ {\rm M}_{\odot}$, the mass-radius relation for dEs branches of from the mass-radius relation defined by nEs and UCDs. In the context of the $\Lambda$CDM-model, this can heuristically, but not quantitatively be understood if the dEs are primordial galaxies, of which some formed the nEs by hierarchical merging (cf. \citealt{White1978,Aarseth1980,Kauffmann1993,Springel2005}).

However, in the light of  the arguments given in Sections~(\ref{sec:dEsTDGs}) and~(\ref{sec:mrUCDsnEs}), it seems to be more likely that the nEs are primordial objects that formed through the monolithic collapse of gas clouds (see Section~\ref{sec:mrUCDsnEs}), while the dEs are secondary objects that formed as TDGs (see Section~\ref{sec:dEsTDGs}). The difference between the mass-radius relation for dEs and the mass-radius relation for nEs would then nevertheless indicate the transition between a primordial and a secondary population of galaxies.

\section{Conclusion}
\label{sec:Conclusion}

\subsection{The nature of old pressure-supported stellar systems}
\label{sec:Conclusion1}

In this paper, the largest existing catalogue of young TDG-candidates is collated and their relation to old pressure-supported stellar systems is discussed. The old stellar systems can be categorized into three groups:

\begin{itemize}
\item dEs, which follow a mass-radius relation quantified by equation~(\ref{eq:dE}) for $10^4 \ {\rm M}_{\odot} < M_{*} < 3\times 10^9 \ {\rm M}_{\odot}$, with a steepening for very low-mass dEs. The properties of these galaxies are best explained with them being ancient TDGs. The reason is that there is plenty of evidence for the existence of young TDGs, both observational (e.g. \citealt{Zwicky1956,Mirabel1992,Duc2011,Miralles2012}) and theoretical (e.g. \citealt{Barnes1992a,Bournaud2006}). These systems would naturally evolve onto the mass-radius sequence defined by the dEs if they survive for a long enough time (see Sections~\ref{sec:evoTDG} and \ref{sec:tidal}) and \citet{Okazaki2000} have shown that already a rather low production rate of TDGs per galaxy encounter would be sufficient to account for all observed dEs (see Section~\ref{sec:dEsTDGs}). Note that TDGs cannot contain a significant amount of CDM \citep{Barnes1992b,Bournaud2010,Kroupa2012}. Consequently, this would also be true for the dEs if they are TDGs. Dynamical $M/L$-ratios derived from spectroscopic measurements indeed suggest that there is little or no CDM in the central parts of dEs with stellar masses $M_{*}\apprge 10^8 \ {\rm M}_{\odot}$ \citep{Forbes2011b,Toloba2011}. At the present, no similar claim can be made for the outskirts of dEs due to the lack of suitable data. The $M/L$-ratios of dEs with lower masses are much higher in many cases, but this does not necessarily indicate that these galaxies are dominated by CDM. The extreme $M/L$-ratios of these galaxies may also indicate that the assumption of virial equilibrium is not valid for them \citep{Kroupa1997,McGaugh2010,Casas2012} or that the laws of gravity have to be modified in the limit of weak gravitational fields \citep{Hernandez2010,McGaugh2010,Kroupa2010,Famaey2011,Kroupa2012}.

\item nEs, which follow a mass-radius relation quantified by equation~(\ref{eq:nE}) for $M_{*} > 3\times 10^9 \ {\rm M}_{\odot}$. Surprisingly, the UCDs lie along the same mass-radius relation (see Section~\ref{sec:oldsystems}), which suggests that the structure of nEs and UCDs was shaped by the same process. This process may be the formation of stellar systems by monolithic collapse of gas clouds (see Section~\ref{sec:mrUCDsnEs}), since this process can explain the mass-radius relation for UCDs \citep{Murray2009}. This would make nEs primordial galaxies. The rapid formation that monolithic collapse implies for the nEs is consistent with the chemical properties of the nEs, namely their large alpha-element enrichment \citep{Thomas2005,Recchi2009}. Also the trend of the nEs being older than the dEs \citep{Cowie1996,Gavazzi2002,Thomas2005,Recchi2009} can easily be explained if nEs are indeed primordial objects and dEs are old TDGs, which can only form after a population of primordial galaxies has formed already. This finding is much more difficult to understand if nEs are built up from dEs via hierarchical merging.

\item GCs and UCDs, which lie along a continuous mass-radius sequence that changes its slope at a mass $M_{*} \approx 2\times 10^{6} \ {\rm M}_{\odot}$ (see, e.g., \citealt{Mieske2008}). GCs and UCDs might well be the same kind of object (e.g. \citealt{Mieske2012}), which would also explain why the formation of GCs and UCDs alike seems to be connected to the interaction between gas-rich galaxies (see \citealt{Zepf1993} for GCs and, e.g., \citealt{Fellhauer2002} for UCDs). This would make GCs and UCDs similar to the TDG-candidates in Tables~(\ref{tab:TDG}) and~(\ref{tab:TDGsim}) as well, and thus to probable progenitors of dEs (see Figure~\ref{fig:MassRadius17}). GCs and UCDs are however much more compact than dEs (see Figure~\ref{fig:MassRadius9}) and their probable progenitors listed in Tables~(\ref{tab:TDG}) and~(\ref{tab:TDGsim}). This indicates a fundamental difference between GCs and UCDs on the one hand and dEs on the other hand (cf. \citealt{Gilmore2007,Misgeld2011}), even if all of them owe their existence to the interactions between galaxies. Interestingly, a difference between likely progenitors of GCs and extended TDGs has apparently been reproduced by \citet{Bournaud2008} in a numerical calculation of the interaction between gas-rich galaxies. It is however still not understood how the different physical processes implemented in the calculation by \citet{Bournaud2008} actually lead to the formation of two distinct types of objects from the matter in tidal tails. 
\end{itemize}

In effect, the observational evidence suggests that all kinds of old pressure-supported stellar systems do not contain CDM.

\subsection{Implications for cosmology}
\label{sec:Conclusion2}

Currently, there are two competing schools of thought in cosmology, i.e. the attempt to describe the Universe and its evolution as a whole. These schools of thought are best distinguished by the conclusions they draw from the fact that general relativity (GR) cannot explain the dynamics of galaxies, if only their visible, baryonic matter is taken into account.

\begin{itemize}
\item According to the first (and at the present dominant) school of thought, GR is an exact formulation of the laws of gravity on all size scales and mass scales. The fact that the dynamics of most galaxies cannot be explained with GR from their baryonic matter would then indicate the presence of unseen, non-baryonic matter in these galaxies. Extensions of the standard model of particle physics predict particles that would be candidates for this kind of dark matter, but experiments with the aim to detect such particles have not been successful so far. The $\Lambda$CDM-model is nevertheless widely accepted, because GR has passed many experimental tests, and because the $\Lambda$CDM-model is a good description of the Universe on large scales (but see \citealt{Kroupa2012}).

\item According to the second school of thought, the dynamics of galaxies is not evidence for the presence of DM in them, but indicates that GR has to be modified in the limit of very weak space-time curvature. This approach has indeed been extremely successful in describing the properties of galaxies (see \citealt{Famaey2011} for a rewiev).
\end{itemize}

 A prediction from the $\Lambda$CDM-model is the 'Dual Dwarf Galaxy Theorem', which states the coexistence of primordial dwarf galaxies and TDGs at masses $M_{*} \apprle10^{10} \ {\rm M}_{\odot}$ \citep{Kroupa2010,Kroupa2012}. The primordial dwarf galaxies would have formed within CDM-haloes, while the TDGs cannot contain CDM. Thus, the primordial dwarf galaxies and the TDGs would have very different matter compositions, which strongly suggests that they should fall into two easily distinguishable groups. Two groups of objects in the appropriate mass-range can indeed be identified in Figure~(\ref{fig:MassRadius9}), namely the UCDs and the dEs. However, according to the conclusions presented in Section~(\ref{sec:Conclusion1}), neither the dEs nor the UCDs seem to be populations of primordial dwarf galaxies within CDM-haloes, but rather populations of objects whose formation was triggered by the tidal interaction between gas-rich galaxies. Thus, the conclusions presented in Section~(\ref{sec:Conclusion1}) support the second school of thought, according to which the $\Lambda$CDM-model needs to be replaced by a cosmological model that is based on a new theory of gravity.

Note however that the data on TDG-candidates used here (see Tables~\ref{tab:TDG} and \ref{tab:TDGsim}) is gathered from different previous publications and is thus based on observations with different instruments, and different numerical calculations, respectively. Moreover, the methods by which the listed parameters have been estimated are in some cases rather crude (cf. Sections~\ref{sec:obsTDG} and~\ref{sec:numTDG}). In order to put our conclusions on a stronger footing, it would be advisable to re-evaluate the existent raw data on TDG-candidates in an effort to make the data as comparable as possible, or even to make a new observational survey of the TDG-candidates.

\section*{Acknowlegdements}
The authors thank R. C. Br\"{u}ns for helpful comments. JD aknowledges support through DFG grant KR1635/13 and a grant for PhD-students from the University of Bonn.

\bibliographystyle{mn2e}
\bibliography{TDGradii}

\newpage

\appendix

\onecolumn

\section[]{Observed Tidal Dwarf Galaxies}
\label{Appendix1}

A list of the data on observed TDG-candidates, as described in Section~(\ref{sec:obsTDG}), is given in Table~(\ref{tab:TDG}).

\begin{center}
\begin{longtable}{lllllllll}
\captionsetup{width=17.65cm}
\caption{\small Data on observed TDG-candidates. Listed are the identification of the object as in the source paper, its half-light radius ($r_{\rm e}$), its equivalent radius ($r$; cf. equation~\ref{eq:equradius}), the mass of its stellar population estimated from its optical luminosity ($M_{*}$), the mass of its stellar population estimated from its H$\alpha$-emission lines ($M_{\alpha}$), the mass of its stellar population under the assumption that the stellar population is a mix of old and young stars ($M_{\rm old}$), the mass of the object estimated from its internal dynamics ($M_{\rm dyn}$), the age of its stellar population assuming a single star burst and finally the reference to the source of the data (1: \citealt{Hunsberger1996}; 2: \citealt{Tran2003}; 3: \citealt{Bournaud2007}; 4: \citealt{Duc2007}; 5: \citealt{Yoshida2008}; 6: \citealt{Galianni2010}; 7: \citealt{Miralles2012}). If the value for $r_{\rm e}$ is given in brackets, it has not been given in the literature, but was calculated here using equations~(\ref{eq:equradius}) and~(\ref{eq:re-estimator}).} \\
\hline
identification				& $r_{\rm e}$ & $r$ 	& $M_{*}$ 		& $M_{\rm H\alpha}$ & $M_{\rm old}$	& $M_{\rm dyn}$ 		& $t$ 			& source \\
						& [pc]	& [pc]	& [${\rm M}_{\odot}$]& [${\rm M}_{\odot}$]	 & [${\rm M}_{\odot}$] & [${\rm M}_{\odot}$]	& years			& \\
\hline
HCG 01b 1				& (1283)	& 4490	& $1.6\times 10^8$	& $-$			& $-$			& $-$				& some $10^8$	& 1 \\
HCG 01b 2				& (563)	& 1970	& $2.5\times 10^7$	& $-$			& $-$			& $-$				& some $10^8$	& 1 \\
HCG 01b 3				& (1180)	& 4130	& $6.3\times 10^8$	& $-$			& $-$			& $-$				& some $10^8$	& 1 \\
HCG 16a 1				& (271)	& 950	& $7.9\times 10^6$	& $-$			& $-$			& $-$				& some $10^8$	& 1 \\
HCG 16a 2				& (286)	& 1000	& $3.2\times 10^6$	& $-$			& $-$			& $-$				& some $10^8$	& 1 \\
HCG 16a 3				& (163)	& 570	& $2.0\times 10^6$	& $-$			& $-$			& $-$				& some $10^8$	& 1 \\
HCG 26b 1				& (962)	& 3370	& $1.6\times 10^8$	& $-$			& $-$			& $-$				& some $10^8$	& 1 \\
HCG 26b 2				& (797)	& 2790	& $6.3\times 10^7$	& $-$			& $-$			& $-$				& some $10^8$	& 1 \\
HCG 26b 3				& (823)	& 2880	& $3.2\times 10^7$	& $-$			& $-$			& $-$				& some $10^8$	& 1 \\
HCG 31a N 1				& (389)	& 1360	& $5.0\times 10^7$	& $-$			& $-$			& $-$				& some $10^8$	& 1 \\
HCG 31a N 2				& (351)	& 1230	& $5.0\times 10^7$	& $-$			& $-$			& $-$				& some $10^8$	& 1 \\
HCG 31a N 3				& (189)	& 660	& $1.3\times 10^7$	& $-$			& $-$			& $-$				& some $10^8$	& 1 \\
HCG 31c N				& (180)	& 630	& $4.0\times 10^7$	& $-$			& $-$			& $-$				& some $10^8$	& 1 \\
HCG 31a S 1				& (169)	& 590	& $7.9\times 10^6$	& $-$			& $-$			& $-$				& some $10^8$	& 1 \\
HCG 31a S 2				& (371)	& 1300	& $1.3\times 10^8$	& $-$			& $-$			& $-$				& some $10^8$	& 1 \\
HCG 38b N				& (591)	& 2070	& $4.0\times 10^7$	& $-$			& $-$			& $-$				& some $10^8$	& 1 \\
HCG 38b S				& (429)	& 1500	& $7.9\times 10^7$	& $-$			& $-$			& $-$				& some $10^8$	& 1 \\
HCG 92c 1				& (474)	& 1660	& $6.3\times 10^6$	& $-$			& $-$			& $-$				& some $10^8$	& 1 \\
HCG 92c 2				& (429)	& 1500	& $7.9\times 10^6$	& $-$			& $-$			& $-$				& some $10^8$	& 1 \\
HCG 92c 3				& (289)	& 1010	& $5.0\times 10^6$	& $-$			& $-$			& $-$				& some $10^8$	& 1 \\
HCG 92c 4				& (557)	& 1950	& $1.3\times 10^7$	& $-$			& $-$			& $-$				& some $10^8$	& 1 \\
HCG 92c 5				& (409)	& 1430	& $1.6\times 10^7$	& $-$			& $-$			& $-$				& some $10^8$	& 1 \\
HCG 92c 6				& (263)	& 920	& $6.3\times 10^6$	& $-$			& $-$			& $-$				& some $10^8$	& 1 \\
HCG 92c 7				& (466)	& 1630	& $1.0\times 10^7$	& $-$			& $-$			& $-$				& some $10^8$	& 1 \\
HCG 92c 8				& (451)	& 1580	& $2.0\times 10^7$	& $-$			& $-$			& $-$				& some $10^8$	& 1 \\
HCG 92c 9				& (517)	& 1810	& $1.3\times 10^7$	& $-$			& $-$			& $-$				& some $10^8$	& 1 \\
HCG 92c 10				& (429)	& 1500	& $2.0\times 10^7$	& $-$			& $-$			& $-$				& some $10^8$	& 1 \\
HCG 92c 11				& (451)	& 1580	& $7.9\times 10^6$	& $-$			& $-$			& $-$				& some $10^8$	& 1 \\
HCG 92c 12				& (591)	& 2070	& $2.5\times 10^7$	& $-$			& $-$			& $-$				& some $10^8$	& 1 \\
HCG 92c 13				& (603)	& 2110	& $5.0\times 10^7$	& $-$			& $-$			& $-$				& some $10^8$	& 1 \\
HCG 92b S 1				& (403)	& 1410	& $4.0\times 10^7$	& $-$			& $-$			& $-$				& some $10^8$	& 1 \\
HCG 92b S 2				& (280)	& 980	& $2.0\times 10^7$	& $-$			& $-$			& $-$				& some $10^8$	& 1 \\
HCG 92b S 3				& (434)	& 1520	& $6.3\times 10^7$	& $-$			& $-$			& $-$				& some $10^8$	& 1 \\
HCG 92b S 4				& (569)	& 1990	& $1.3\times 10^8$	& $-$			& $-$			& $-$				& some $10^8$	& 1 \\
HCG 92b S 5				& (377)	& 1320	& $2.0\times 10^7$	& $-$			& $-$			& $-$				& some $10^8$	& 1 \\
HCG 92b N 1				& (497)	& 1740	& $2.0\times 10^7$	& $-$			& $-$			& $-$				& some $10^8$	& 1 \\
HCG 92b N 2				& (534)	& 1870	& $1.6\times 10^8$	& $-$			& $-$			& $-$				& some $10^8$	& 1 \\
HCG 92b N 3				& (406)	& 1420	& $2.5\times 10^7$	& $-$			& $-$			& $-$				& some $10^8$	& 1 \\
HCG 92d S 1				& (417)	& 1460	& $5.0\times 10^7$	& $-$			& $-$			& $-$				& some $10^8$	& 1 \\
HCG 92d S 2				& (274)	& 960	& $2.5\times 10^7$	& $-$			& $-$			& $-$				& some $10^8$	& 1 \\
HCG 92d N 1				& (429)	& 1500	& $2.5\times 10^7$	& $-$			& $-$			& $-$				& some $10^8$	& 1 \\
HCG 92d N 2				& (249)	& 870	& $1.0\times 10^7$	& $-$			& $-$			& $-$				& some $10^8$	& 1 \\
HCG 92d N 3				& (300)	& 1050	& $4.0\times 10^7$	& $-$			& $-$			& $-$				& some $10^8$	& 1 \\
HCG 92d N 4				& (440)	& 1540	& $2.5\times 10^7$	& $-$			& $-$			& $-$				& some $10^8$	& 1 \\
HCG 96c E				& (674)	& 2360	& $5.0\times 10^7$	& $-$			& $-$			& $-$				& some $10^8$	& 1 \\
HCG 96a W 1				& (729)	& 2550	& $3.2\times 10^7$	& $-$			& $-$			& $-$				& some $10^8$	& 1 \\
HCG 96a W 2				& (631)	& 2210	& $6.3\times 10^7$	& $-$			& $-$			& $-$				& some $10^8$	& 1 \\
UGC 10214 SSC			& 161	& $-$	& $6.6\times 10^5$	& $-$			& $-$			& $-$				& $4-5\times 10^6$	& 2 \\
NGC5291N				& (1057)	& 3700	& $1.1\times 10^8$	& $-$			& $-$			& $3.0\times10^9$		& $<5\times 10^6$	& 3 \\
NGC5291S				& (1429)	& 5000	& $7.5\times 10^7$	& $-$			& $-$			& $2.7\times10^9$		& $<5\times 10^6$	& 3 \\
NGC5291SW				& (571)	& 2000	& $3.0\times 10^7$	& $-$			& $-$			& $1.2\times10^9$		& $<5\times 10^6$	& 3 \\
VCC 2062				& (600)	& 2100	& $5.0\times 10^7$	& $-$			& $-$			& $3.5\times10^8$		& $3\times 10^8$	& 4 \\
RB 199 Knot 1				& 250	& $-$	& $8.8\times 10^7$	& $-$			& $-$			& $-$				& some $10^8$	& 5 \\
RB 199 Knot 2				& 250	& $-$	& $1.4\times 10^8$	& $-$			& $-$			& $-$				& some $10^8$	& 5 \\
RB 199 Knot 3				& 250	& $-$	& $4.8\times 10^7$	& $-$			& $-$			& $-$				& some $10^8$	& 5 \\
RB 199 Knot 4				& 250	& $-$	& $7.6\times 10^6$	& $-$			& $-$			& $-$				& some $10^8$	& 5 \\
RB 199 Knot 5				& 250	& $-$	& $2.3\times 10^7$	& $-$			& $-$			& $-$				& some $10^8$	& 5 \\
RB 199 Knot 6				& 250	& $-$	& $2.0\times 10^7$	& $-$			& $-$			& $-$				& some $10^8$	& 5 \\
NGC 1097 Knot A			& 336	& $-$	& $6.0\times 10^6$	& $-$			& $-$			& $-$				& some $10^9$	& 6 \\
NGC 1097 Knot B			& 482	& $-$	& $4.0\times 10^6$	& $-$			& $-$			& $-$				& some $10^9$	& 6 \\
IRAS 04315$-$0840 1		& 38		& 166	& $7.9\times 10^4$	& $2.5\times 10^5$	& $6.3\times 10^5$	& $3.2\times 10^7$		& $4.6\times 10^6$	& 7 \\
IRAS 04315$-$0840 2		& 21		& 93		& $1.6\times 10^4$	& $6.3\times 10^5$	& $3.2\times 10^5$	& $6.3\times 10^7$		& $7.0\times 10^6$	& 7 \\
IRAS 06076$-$2139 1		& 77		& 306	& $2.5\times 10^5$	& $7.9\times 10^5$	& $-$			& $1.0\times 10^8$		& $4.2\times 10^6$	& 7 \\
IRAS 06076$-$2139 2		& 59		& 283	& $7.9\times 10^4$	& $2.5\times 10^5$	& $-$			& $1.0\times 10^8$		& $4.5\times 10^6$	& 7 \\
IRAS 06076$-$2139 3		& 53		& 121	& $4.0\times 10^4$	& $1.3\times 10^5$	& $4.0\times 10^5$	& $7.9\times 10^7$		& $5.4\times 10^6$	& 7 \\
IRAS 06076$-$2139 4		& 51		& 116	& $3.2\times 10^4$	& $1.0\times 10^5$	& $-$			& $1.0\times 10^8$		& $4.8\times 10^6$	& 7 \\
IRAS 06076$-$2139 5		& 41		& 89		& $2.5\times 10^4$	& $1.3\times 10^5$	& $-$			& $4.0\times 10^7$		& $4.9\times 10^6$	& 7 \\
IRAS 06076$-$2139 6		& 66		& 137	& $1.6\times 10^5$	& $7.9\times 10^5$	& $-$			& $-$				& $4.9\times 10^6$	& 7 \\
IRAS 07027$-$6011 S 1		& 31		& 127	& $7.9\times 10^4$	& $4.0\times 10^5$	& $-$			& $7.9\times 10^7$		& $3.6\times 10^6$	& 7 \\
IRAS 07027$-$6011 S 2 		& 61		& 186	& $3.2\times 10^4$	& $4.0\times 10^4$	& $-$			& $-$				& $3.2\times 10^6$	& 7 \\
IRAS 08572$+$3915 N		& 105	& 405	& $2.5\times 10^5$	& $7.9\times 10^5$	& $-$			& $5.0\times 10^8$		& $4.0\times 10^6$	& 7 \\
IRAS 08572$+$3915 SE 3	& 76		& 322	& $1.3\times 10^5$	& $1.3\times 10^5$	& $1.3\times 10^6$	& $-$				& $4.9\times 10^6$	& 7 \\
IRAS 08572$+$3915 SE 4	& 191	& 312	& $3.2\times 10^5$	& $1.3\times 10^6$	& $3.2\times 10^6$	& $-$				& $5.9\times 10^6$	& 7 \\
IRAS F10038$-$3338 3  		& 43		& 269	& $2.5\times 10^4$	& $3.2\times 10^4$	& $5.0\times 10^5$	& $4.0\times 10^7$		& $3.6\times 10^6$	& 7 \\
IRAS F10038$-$3338 4  		& 88		& 252	& $1.0\times 10^5$	& $1.6\times 10^5$	& $1.0\times 10^6$	& $1.6\times 10^8$		& $4.9\times 10^6$	& 7 \\
IRAS 12112$+$0305 1   		& 200	& 887	& $1.3\times 10^7$	& $2.0\times 10^7$	& $1.0\times 10^8$	& $2.5\times 10^9$		& $4.3\times 10^6$	& 7 \\
IRAS 12112$+$0305 4   		& 82		& 288	& $2.5\times 10^5$	& $1.3\times 10^6$	& $-$			& $-$				& $4.2\times 10^6$	& 7 \\
IRAS 14348$-$1447 1   		& 280	& 909	& $5.0\times 10^7$	& $1.3\times 10^8$	& $3.2\times 10^8$	& $2.0\times 10^9$		& $4.4\times 10^6$	& 7 \\
IRAS 15250$+$3609 1   		& 165	& 627	& $4.0\times 10^6$	& $5.0\times 10^6$	& $3.2\times 10^7$	& $2.0\times 10^9$		& $5.3\times 10^6$	& 7 \\
IRAS F18093$-$5744 N  		& 20		& 74		& $1.0\times 10^5$	& $6.3\times 10^5$	& $-$			& $7.9\times 10^7$		& $5.1\times 10^6$	& 7 \\
IRAS F18093$-$5744 C  		& 37		& 78		& $6.3\times 10^4$	& $1.6\times 10^5$	& $5.0\times 10^5$	& $4.0\times 10^7$		& $4.9\times 10^6$	& 7 \\
IRAS 23128$-$5919     		& 83		& 376	& $5.0\times 10^5$	& $1.3\times 10^6$	& $3.2\times 10^6$	& $1.0\times 10^9$		& $4.8\times 10^6$	& 7 \\
IRAS 16007$+$3743 R1  		& 828	& $-$	& $6.3\times 10^8$	& $-$			& $-$			& $6.3\times 10^9$		& $7.1\times 10^6$	& 7 \\
IRAS 16007$+$3743 R2  		& 884	& $-$	& $6.3\times 10^8$	& $-$			& $-$			& $1.0\times 10^{10}$	& $5.4\times 10^6$	& 7 \\
IRAS 16007$+$3743 R3  		& 851	& $-$	& $6.3\times 10^7$	& $-$			& $-$			& $1.3\times 10^{10}$	& $6.4\times 10^6$	& 7 \\
\hline
\label{tab:TDG}
\end{longtable}
\end{center}

\clearpage

\section[]{Numerical calulations on the formation of Tidal Dwarf Galaxies}
\label{Appendix2}

A list of the data on TDGs that formed in calculations of the encounters between gas-rich galaxies, as described in Section~(\ref{sec:numTDG}), is given in Table~(\ref{tab:TDGsim}).

\begin{table}
\caption{Data on TDGs that were found in numerical calculations of encounters between gas-rich galaxies. Listed are the effective radius of each TDG ($r_{\rm e}$), and if available its size (given through diameters along two orthogonal axes), the mass of its stellar population ($M_{*}$), its total mass ($M$), the time at the end of the calculation ($t$) and finally the reference to the source of the data (1: \citealt{Bournaud2008}; 2: \citealt{Wetzstein2007}; 3: \citealt{Barnes1992a}). The value of $M_{*}$ for the TDG from \citet{Barnes1992a} is an estimate based on $M$.}
\centering
\begin{tabular}{llllll}
\hline
$r_{\rm e}$	& size				& $M_{*}$				& $M$			& $t$				& source \\
$[{\rm pc}]$			& [pc] $\times$ [pc]		& $[{\rm M}_{\odot}]$		& $[{\rm M}_{\odot}]$ & years			& \\
\hline
8.5  			& $61\times 46 $		& $6.8\times 10^6$		& $-$			& $9.5\times 10^8$	& 1 \\
7.6  	 		& $63\times 35 $		& $8.7\times 10^6$		& $-$			& $9.5\times 10^8$	& 1 \\
9.7  			& $46\times 78 $		& $1.9\times 10^7$		& $-$			& $9.5\times 10^8$	& 1 \\
460  			& $3700\times 2200 $	& $2.7\times 10^8$		& $-$			& $9.5\times 10^8$	& 1 \\
420  			& $4500\times 1500 $	& $5.2\times 10^8$		& $-$			& $9.5\times 10^8$	& 1 \\
700  			& $-$ 				& $1.0\times 10^8$		& $3.5\times 10^8$	& $1.2\times 10^8$	& 2 \\
229 			& $-$				& $(1.3\times 10^8)$		& $4.0\times 10^8$	& $7.5\times 10^8$	& 3 \\
\hline
\end{tabular}
\label{tab:TDGsim}
\end{table}

\label{lastpage}

\end{document}